\documentclass[conference]{IEEEtran}
\IEEEoverridecommandlockouts
\usepackage{cite}
\usepackage{amsmath,amssymb,amsfonts,latexsym}
\usepackage{amsthm}
\usepackage{braket}
\usepackage{graphicx}
\usepackage{textcomp}
\usepackage{xcolor}
\usepackage{mathtools}
\usepackage{multirow}
\usepackage{url}
\usepackage{booktabs}
\usepackage{hyperref}
\usepackage{lipsum}
\usepackage{nicematrix}
\usepackage[all]{xy}
\usepackage{tikz-cd}
\usepackage{tikz}
\usepackage{makecell}
\usepackage{caption}
\usepackage{subcaption}
\usepackage{quantikz}

\usetikzlibrary{calc}
\usetikzlibrary{intersections,shapes.arrows}
\usetikzlibrary{arrows.meta, positioning, backgrounds, decorations.markings}
\usetikzlibrary{arrows.meta,decorations.markings,calc}

\usepgfmodule{nonlineartransformations}

\def \a{\alpha}

\newtheorem{theorem}{Theorem}
\newtheorem*{theorem*}{Theorem}
\newtheorem*{lemma*}{Lemma}
\newtheorem*{example*}{Example}



\def\BibTeX{{\rm B\kern-.05em{\sc i\kern-.025em b}\kern-.08em
    T\kern-.1667em\lower.7ex\hbox{E}\kern-.125emX}}

\makeatletter
\newcommand{\linebreakand}{%
  \end{@IEEEauthorhalign}
  \hfill\mbox{}\par
  \mbox{}\hfill\begin{@IEEEauthorhalign}
}
\makeatother

\begin{document}

\title{Quasi-polar Decomposition of Quantum Neural Networks via Adaptive Non-local Observables\thanks{
The views expressed in this article are those of the authors and do not represent the views of Wells Fargo. This article is for informational purposes only. Nothing contained in this article should be construed as investment advice. Wells Fargo makes no express or implied warranties and expressly disclaims all legal, tax, and accounting implications related to this article.}}

\author{Shih-Hao Ho$^1$, Yan Li$^2$, Huan-Hsin Tseng$^3$, Hsin-Yi Lin$^3$, Samuel Yen-Chi Chen$^4$, Shinjae Yoo$^3$\\
\small $^1$Independent researcher, Taiwan\\
\small $^2$Department of Electrical Engineering, The Pennsylvania State University, University Park, PA, USA\\
\small $^3$AI \& ML Department, Brookhaven National Laboratory, Upton NY, USA\\
\small $^4$Wells Fargo, New York NY, USA\\
\small $^1$\texttt{shihhao.ho@gmail.com}\\
\small $^2$\texttt{yql5925@psu.edu}\\
\small $^3$\texttt{\{htseng, hlin7, sjyoo\}@bnl.gov}\\
\small $^4$\texttt{ycchen1989@ieee.org}
}

\maketitle

\begin{abstract}
We use Diagonal Adaptive Non-local Observables (DANO) as a canonical decomposition for studying Variational Quantum Circuit model evolution. Separating each learned observable into a diagonal spectrum and a unitary basis gives a quasi-polar description: the spectral weights are viewed as radial coordinates, while the unitary circuit serves as angular coordinates through Lie group identifications. This turns the training process into a trajectory in spectral and Lie-algebra space.

Experiments on two classification tasks show that DANO radial spectral expansion correlates with accuracy. DANO angle coordinates reveal a dominant accuracy-correlated component. The framework provides a different perspective to characterize quantum model behavior.
\end{abstract}


\section{Introduction}\label{sec:Indroduction}

Variational quantum algorithms (VQAs) optimize parameterized quantum circuits through expectation values of observables~\cite{cerezo2021variational}. Quantum machine learning extends this circuit-measurement structure to data-driven tasks~\cite{biamonte2017quantum}. Existing interpretive viewpoints often focus on the input and circuit side: quantum feature maps define kernel geometries~\cite{havlicek2019supervised}, data-dependent analyses test when quantum models can differ from classical learners~\cite{huang2021power}, and Fourier analyses show how encoding gates determine accessible frequency spectra~\cite{schuld2021dataencoding}. Lie-algebraic approaches further relate circuit generators to trainability and expressivity~\cite{ragone2024lie}.

This paper instead focuses on the observable side of the circuit. Quantum convolutional neural networks and adaptive measurement schemes suggest that measurement design can be part of the model architecture~\cite{cong2019quantum,garciaPerez2021learning}. Adaptive Non-local Observables (ANO) make this explicit by considering dynamical Hermitian operators~\cite{ANO}. Diagonal Adaptive Non-local Observables (DANO)~\cite{tseng2026diagonal} further keep the spectral weights explicit while using a circuit to supply the measurement basis.

Using the DANO factorization, the quantum model training process may be viewed through the DANO quasi-polar decomposition. Spectral weights play the radial role, while unitary basis rotations play the angular role. This provides a tool for analyzing how quantum models evolve during optimization.

\section{Background}\label{Sec: Background}

Denote an $n$-qubit system with Hilbert space $\mathcal{H}^n:=(\mathbb{C}^2)^{\otimes n}$ and unitary group $\mathcal{U}(\mathcal{H}^n):=\{ U \in \operatorname{End}(\mathcal{H}^n) \, | \, U^\dagger U=UU^\dagger=I_{2^n}\}$
with $\mathbb{H}(n):=\{A\in\operatorname{End}(\mathcal{H}^n) \, | \, A = A^\dagger \}$.  %

Let $\mathcal{X}\subset\mathbb{R}^n$ be the input space and $x = (x_1, \ldots, x_n) \in\mathcal{X}$ a classical input. Denote $\ket{\psi_0}\in\mathcal{H}^n$ an initial state.  A Variational Quantum Circuit (VQC) consists of an encoding $V:\mathcal{X} \to \mathcal{U}(\mathcal{H}^n)$, a trainable unitary $U(\theta) \in \mathcal{U}(\mathcal{H}^n)$ with parameters $\theta \in \mathbb{R}^m$, and a fixed observable $H_0\in\mathbb{H}(n)$.  With $\ket{\psi_x}:=V(x)\ket{\psi_0}$, the VQC output is defined as,
\begin{equation}\label{E: VQC}
    f_{\theta,H_0}(x) :=\bra{\psi_x}U^\dagger(\theta)H_0U(\theta)\ket{\psi_x}.
\end{equation}

\subsection{Adaptive Non-local Observables}\label{Subsec: ANO}

ANO is motivated by denoting $H(\theta) := U^\dagger(\theta) H_0 U(\theta)$, then VQC Eq.~(\ref{E: VQC}) is equivalent to $f_{\theta,H_0}(x)=\bra{\psi_x}H(\theta)\ket{\psi_x}$, where the circuit optimization can now be regarded as a dynamic motion on $\mathbb{H}(n)$~\cite{ANO}. 

More generally, let $Q\subset\{1,\ldots,n\}$ be a $k$-element set of qubit indices and let $P_Q:\mathcal{H}^n\to\mathcal{H}^k\otimes\mathcal{H}^{n-k}$ be the coordinate transformation transposing the qubits of $Q$ to the first $k$ slots. Then an ANO is defined as an observable $\widetilde H(\eta) \in \mathbb{H}(k)$ on $Q$ such that its embedded $n$-qubit observable is
\begin{equation}\label{E: H tensor product}
    H_Q(\eta) := P_Q^\dagger \bigl( \widetilde H(\eta) \otimes I_{2^{n-k}} \bigr)P_Q \in \mathbb{H}(n).
\end{equation}
where $\eta\in\mathbb{R}^{K^2}$ are tunable parameters. The corresponding measurement, after a variational unitary in $\mathcal{U}(\mathcal{H}^n)$, is $\bra{\psi_x}U^\dagger(\theta)H_Q(\eta)U(\theta)\ket{\psi_x}$. Since the space $\mathbb{H}(k)$ has dimension $K^2=4^k$, these become the free parameters of $\eta$.

\subsection{Diagonal Adaptive Non-local Observables}\label{Sec: DANO}

DANO~\cite{tseng2026diagonal} is a canonical case of ANO, where only the spectral variables are changed. By the spectral theorem, every $\widetilde H\in\mathbb{H}(k)$ admits,
\begin{equation}\label{E: diag H}
    \widetilde H=U^\dagger\Lambda(\lambda)U, \quad
    \Lambda(\lambda):=\operatorname{diag}(\lambda_1,\ldots,\lambda_K),
\end{equation}
for some $U \in \mathcal{U}(\mathcal{H}^k)$ and $\lambda = (\lambda_1, \ldots, \lambda_K) \in \mathbb{R}^K$. Denote the unitaries generated by the chosen ansatz $\mathcal{U}_{\rm ans}\subseteq\mathcal{U}(\mathcal{H}^k)$ and define
\begin{equation}\label{E: DANO}
    \operatorname{DANO}_k(\mathcal{U}_{\rm ans})
    :=\{U^\dagger\Lambda(\lambda)U:\lambda\in\mathbb{R}^{K},\ U\in\mathcal{U}_{\rm ans}\}.
\end{equation}
Thus DANO replaces the $K^2$ parameters of a general ANO block by the $K$ spectral parameters $\lambda$, while the ansatz supplies the measurement basis.  This DANO becomes the foundation we develop below.

\section{``Polar decompositions'' of Adaptive Non-local Observables}\label{Sec: Polar decom}

DANO~(\ref{E: diag H}) provides a geometric picture for viewing quantum models through a \textit{polar-coordinate-like} description of observables. The usual polar coordinates on $\mathbb{C}\simeq\mathbb{R}^2$ have the form
\begin{equation}\label{E: polar coord}
p = re^{i\vartheta}, \quad r\geq 0, \quad \vartheta \in[0,2\pi).
\end{equation}
Comparing this with Eq.~(\ref{E: diag H}), the eigenvalues $\lambda$ are analogous to the radial part, while the unitary $U$ plays the role of an angular coordinate. This analogy is most transparent when $\Lambda(\lambda)\succeq 0$ (positive semi-definite), in parallel with the condition $r\geq 0$. To make the angular analogy more precise, we need a mathematical result from Lie theory.

\begin{theorem}\label{Thm: Unitary group}
Let $\mathcal{U}(K) = \{ W \in M_K(\mathbb{C})\mid W^\dagger W=I_K \}$ and its Lie algebra $\mathfrak{u}(K) = \{ X \in M_K(\mathbb{C}) \mid X^\dagger = -X \}$. The exponential map $\exp:\mathfrak{u}(K)\to \mathcal{U}(K)$ is surjective. That is, for every $U \in \mathcal{U}(K)$ there exists a Hermitian matrix $\Theta = \Theta^\dagger \in M_K(\mathbb{C})$ such that $U = e^{i\Theta}$.
\end{theorem}
Thus, with Theorem~\ref{Thm: Unitary group} DANO Eq.~(\ref{E: DANO}) can be written as
\begin{equation}\label{E: polar Hermitian exponential}
\widetilde H(\lambda,\Theta) = e^{-i\Theta} \, \Lambda(\lambda) \, e^{i\Theta}.
\end{equation}

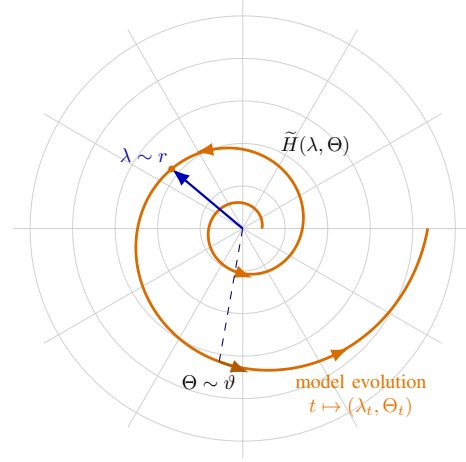
\begin{figure}[t]
    \vskip -0.1in
\centering

\newcommand{\DANOfigscale}{0.75}

\begin{tikzpicture}[
    scale=\DANOfigscale,
    transform shape,
    >=Latex,
    axis/.style={->, line width=0.45pt, black!65},
    gridline/.style={line width=0.25pt, black!18},
    radial/.style={->, line width=0.85pt, blue!70!black},
    angular/.style={->, line width=0.85pt, magenta!75!black},
    spiral/.style={
        line width=1.05pt,
        orange!85!black,
        smooth
    },
    panel/.style={
        fill=white,
        draw=black!10,
        fill opacity=0.97,
        text opacity=1,
        rounded corners=2pt,
        inner sep=2.4pt,
        font=\scriptsize
    },
    minilabel/.style={
        fill=white,
        fill opacity=0.97,
        text opacity=1,
        rounded corners=1.5pt,
        inner sep=1.5pt,
        font=\scriptsize
    }
]

\pgfmathsetmacro{\spiralA}{0.34}
\pgfmathsetmacro{\spiralB}{0.18}
\pgfmathsetmacro{\angmax}{720}
\pgfmathsetmacro{\R}{4.05}

\foreach \rr in {0.75,1.50,2.25,3.00,3.75}
    \draw[gridline] (0,0) circle (\rr);

\foreach \aa in {0,30,...,330}
    \draw[gridline] (0,0) -- ({\R*cos(\aa)},{\R*sin(\aa)});

\draw[spiral, domain=0:\angmax, samples=440, variable=\x]
    plot (
        {\spiralA*exp(\spiralB*pi*\x/180)*cos(\x)},
        {\spiralA*exp(\spiralB*pi*\x/180)*sin(\x)}
    );

\foreach \aa in {260,470,665}{
    \pgfmathsetmacro{\rr}{\spiralA*exp(\spiralB*pi*\aa/180)}
    \pgfmathsetmacro{\xx}{\rr*cos(\aa)}
    \pgfmathsetmacro{\yy}{\rr*sin(\aa)}
    \pgfmathsetmacro{\tx}{\spiralB*cos(\aa)-sin(\aa)}
    \pgfmathsetmacro{\ty}{\spiralB*sin(\aa)+cos(\aa)}
    \draw[->, line width=0.8pt, orange!85!black]
        (\xx,\yy) -- ($(\xx,\yy)+0.32*(\tx,\ty)$);
}

\pgfmathsetmacro{\angP}{500}
\pgfmathsetmacro{\angVis}{140}
\pgfmathsetmacro{\rP}{\spiralA*exp(\spiralB*pi*\angP/180)}
\pgfmathsetmacro{\xP}{\rP*cos(\angP)}
\pgfmathsetmacro{\yP}{\rP*sin(\angP)}
\coordinate (P) at (\xP,\yP);

\draw[radial] (0,0) -- (P);
\fill[orange!85!black] (P) circle (1.7pt);

\node[text=blue!70!black, anchor=east] at (-1.2,1.3)
    {$\lambda \sim r$};

\node[text=black, anchor=west] at (-1.2,-2.7)
    {$\Theta \sim \vartheta$};

\node[text=black, anchor=south west] at ($(P)+(1.8,0.1)$)
    {$\widetilde H(\lambda,\Theta)$};

\pgfmathsetmacro{\angQ}{620}
\pgfmathsetmacro{\rQ}{\spiralA*exp(\spiralB*pi*\angQ/180)}
\pgfmathsetmacro{\xQ}{\rQ*cos(\angQ)}
\pgfmathsetmacro{\yQ}{\rQ*sin(\angQ)}
\coordinate (Q) at (\xQ,\yQ);

\draw[thin, blue!45!black, dashed] (0,0) -- (Q);

\pgfmathsetmacro{\dx}{\spiralB*cos(\angQ)-sin(\angQ)}
\pgfmathsetmacro{\dy}{\spiralB*sin(\angQ)+cos(\angQ)}

\draw[->, line width=0.85pt, orange!70!black]
    (Q) -- ($(Q)+0.55*(\dx,\dy)$);

\node[text=orange!90!black, anchor=north] at ($(Q)+(2.5,-0.1)$)
    {model evolution};

\node[text=orange!90!black, anchor=north] at ($(Q)+(2.5,-0.5)$)
    {$t \mapsto (\lambda_t, \Theta_t)$};

\end{tikzpicture}

\caption{
DANO decomposition gives a quasi-polar description for quantum models $\widetilde H(\lambda,\Theta)=e^{-i\Theta}\Lambda(\lambda)e^{i\Theta} \Leftrightarrow p=re^{i\vartheta} $ with $(\lambda, \Theta) \Leftrightarrow (r, \theta)$. Model evolution yields a trajectory.}

\label{fig:DANO_polor}
    \vskip -0.1in
\end{figure}

Although the Hermitian generator $\Theta=\Theta^\dagger$ is generally not unique, the anti-Hermitian matrix $i\Theta\in\mathfrak{u}(K)$ is the corresponding Lie-algebra element for the circuit configuration $U(\theta)$. From DANO Eq.~(\ref{E: polar Hermitian exponential}), a training model at time (iteration) $t$ creates a trajectory,
\begin{equation}\label{E: radial angular trajectories}
    t \mapsto (\lambda_t, \Theta_t),
\end{equation}
where the history of $\lambda_t$ records spectral motion of the observable, and $\Theta_t$ records the Hermitian-generator representation of the angular motion, equivalently $i\Theta_t\in\mathfrak{u}(K)$ records the associated Lie-algebra direction; see \textbf{\textbf{Fig.}}~\ref{fig:DANO_polor}.

Since Theorem~\ref{Thm: Unitary group} is not a \textit{constructive} statement, below we give an explicit example  to see how a Lie generator $\Theta$ is found.

\begin{example*}[A 2-qubit, 2-layered circuit]\label{Ex: local-to-single-exp}
Let $A_{j\ell} = A_{j\ell}^\dagger\in M_2(\mathbb{C})$, where $j \in \{1,2\}$ indexes qubits and $\ell \in \{1,2\}$ indexes layers.  For a 2-layered (local) circuit
\begin{equation}\label{E: local two layer unitary}
    U = \big(e^{iA_{12}}\otimes e^{iA_{22}}\big)
        \big(e^{iA_{11}}\otimes e^{iA_{21}}\big) := U_1 \otimes U_2,
\end{equation}
then $U_1 = e^{iA_{12}} \, e^{iA_{11}}$ and $U_2 = e^{iA_{22}} \, e^{iA_{21}}$. Choose Hermitian logarithms $B_j = -i \log U_j$, so that $U_j = e^{i B_j}$.  Then
\begin{equation}\label{E: local theta generator}
    \Theta = B_1\otimes I_2 + I_2\otimes B_2
\end{equation}
satisfies $U = e^{i\Theta}$.  By the Baker-Campbell-Hausdorff,
\begin{equation}\label{E: BCH local generators}
\begin{aligned}
    B_1 &= A_{11} + A_{12} + \frac{i}{2}[A_{12},A_{11}]+\cdots,\\
    B_2 &= A_{21} + A_{22} + \frac{i}{2}[A_{22},A_{21}]+\cdots .
\end{aligned}
\end{equation}
Eq.~(\ref{E: local theta generator}) constructs $\Theta$ explicitly as described in Theorem~\ref{Thm: Unitary group}, which also reveals that $\Theta$ depends nontrivially on single-qubit generators $ \{ A_{j \ell} \}$ in general.

\end{example*}

\section{Experiments}\label{sec_exp_results}

We evaluate the quasi-polar coordinates of Eq.~\eqref{E: radial angular trajectories} on two 10-class datasets: 1) \textit{MNIST} and 2) a subset of \textit{Extended Yale Face Database B}~\cite{georghiades2001illuminationCone} (\textbf{\textbf{Fig.}}~\ref{fig: Yale B}).  Inputs are dimensionally reduced by PCA to $x\in[-\pi,\pi]^{10}$ and labels are $\{0, \ldots, 9\}$; both datasets use a stratified $80/10/10$ train/validation/test split. Throughout, we use 10 qubits, 6-local ($k=6$, $K=2^k=64$) for experiments.

\subsection{Model and optimization}\label{subsec:polar_model}

Let $\ket{\psi_x}=V(x)\ket{0}^{\otimes n}$ as in Eq.~\eqref{E: VQC}.  The encoding $V(x)$ is a Hadamard layer followed by $R_y(x_j)$ on qubit $j$. The ansatz $U(\theta)$ has depth $L=6$.  Each layer applies nearest-neighbor CNOTs in even-odd and shifted odd-even pairs, followed by $R_y(\theta_{\ell j})$. Thus, $\theta\in\mathbb{R}^{L\times n}$ has $60$ parameters.

For each $q\in\{1,\ldots,n\}$, define cyclic qubit index set,
\[
    Q_q = (q, q + 1, \ldots, q + k - 1) \pmod n .
\]
The DANO given by Eq.~\eqref{E: H tensor product},~\eqref{E: diag H} with eigenvalues $\lambda_q\in\mathbb{R}^K$ is denoted as $H_q(\lambda_q)$. Then the (output) measurements are,
\[
    z_q(x;\theta, \lambda) := \bra{\psi_x}U^\dagger(\theta)H_q(\lambda_q)U(\theta)\ket{\psi_x}, \,\,
    z=(z_1,\ldots,z_n).
\]

Thus, the trainable parameters are $(\lambda_t,\theta_t)$, with $|\lambda_t|=n\times K=640$ and $|\theta_t|=n\times L=60$ at each time $t$; $\Theta_t = -i\log U(\theta_t)$ is used for analysis. Training is using Adam for 30 epochs and batch size 32, learning rates $10^{-3}$ for $\theta$ and $10^{-1}$ for $\lambda$. We use alternating updates, 5 iterations in $\theta$ with $\lambda$ fixed, followed by 5 steps in $\lambda$ with $\theta$ fixed.

\subsection{Radial and angular decomposition of DANO}\label{subsec:polar_coordinates_exp}

To observe the \textit{evolution of training quantum models} (conceptual \textbf{Fig.}~\ref{fig:DANO_polor}), our quasi-polar mapping Eq.~\eqref{E: polar Hermitian exponential} is applied at each training iteration $t$ with the first 3 largest eigenvalues as radial coordinates $\lambda_{t,q}$ visualized,
\begin{equation}\label{E: top k eigenvalues}
     \rho_{t,q}:=\operatorname{top}_3(\lambda_{t,q}) \in \mathbb{R}^3,
\end{equation}
The angular coordinate $\Theta_t$ at time $t$ is given by Theorem~\ref{Thm: Unitary group},
\begin{equation}\label{E: unitary angle}
    U(\theta_t) = e^{i \Theta_t} \quad ( \Theta_t^\dagger =  \Theta_t)
\end{equation}
The lift uses global-phase correction, polar projection to the nearest unitary when needed, Schur decomposition, and phase unwrapping.

We first use the Extended Yale Face Database B (\textbf{\textbf{Fig.}}~\ref{fig: Yale B}), reduced by PCA to $n=10$ (real) inputs and encoded into a 10-qubit DANO. We observe the following in our quasi-polar coordinates:
\begin{figure}[htbp]
\vskip -0.1in
\begin{center}
\centerline{\includegraphics[width=0.8\columnwidth]{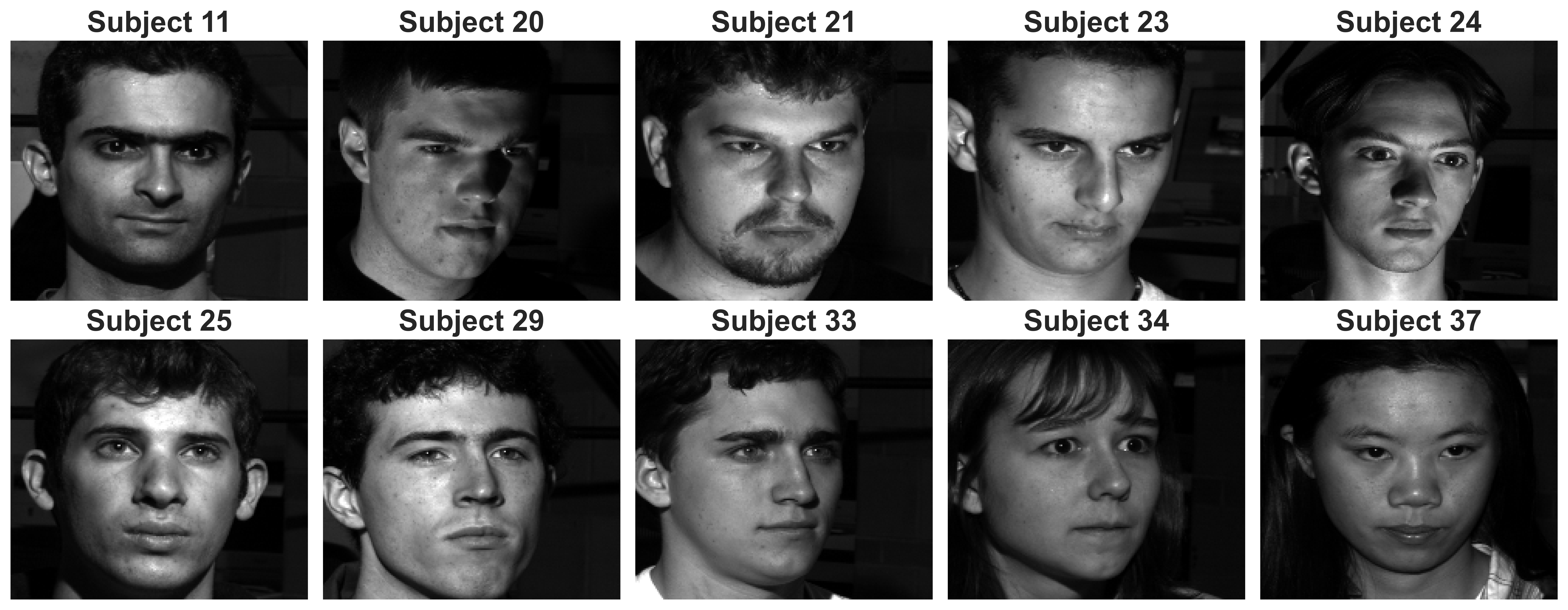}}
\caption{\textbf{Yale B}: 10 individuals selected for classification.}
\label{fig: Yale B}
\end{center}
\vskip -0.15in
\end{figure}

\begin{figure}[htbp]
    \vskip -0.15in
\centering
\includegraphics[width=0.7\columnwidth]{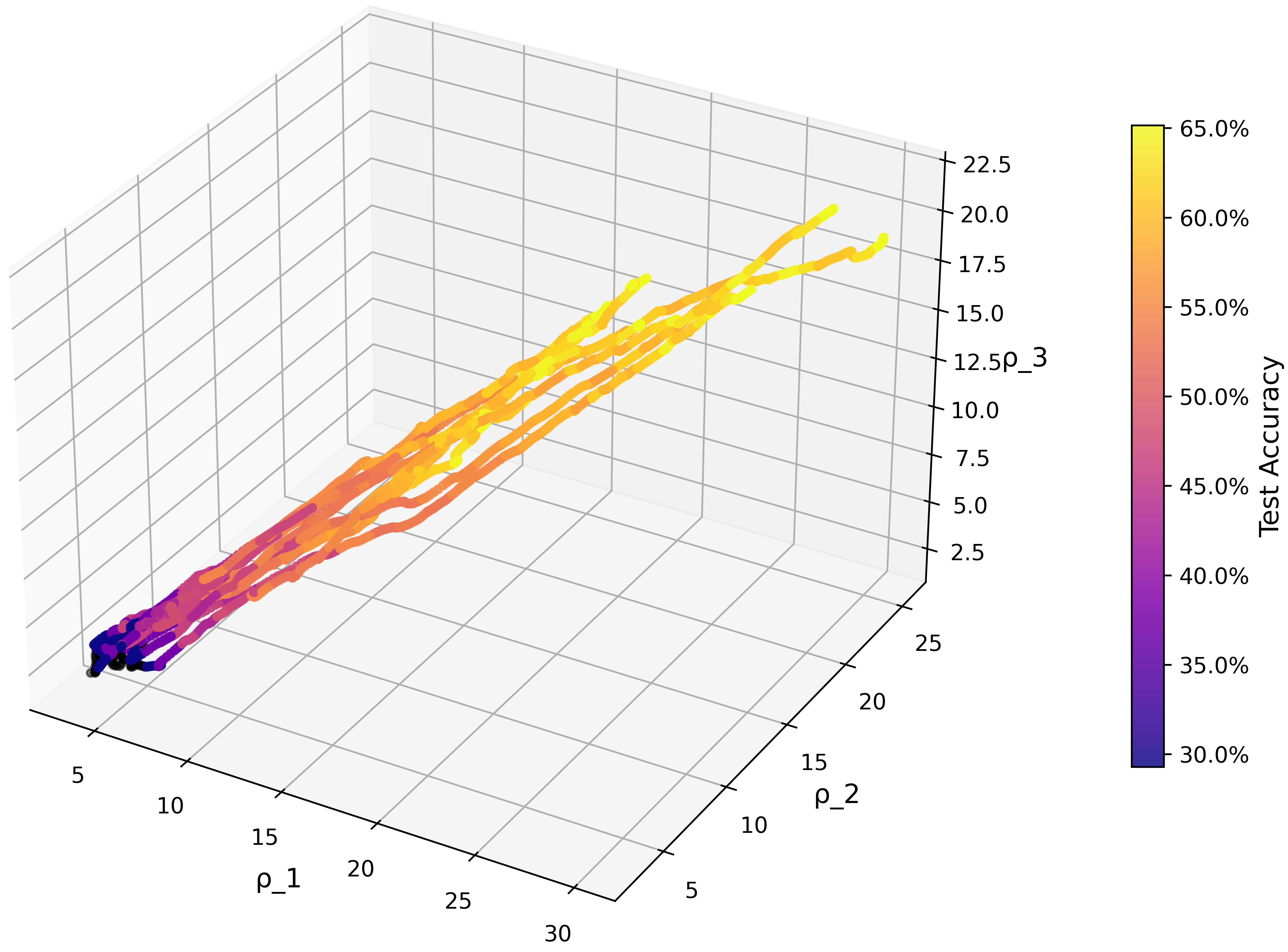}
\caption{\textbf{Yale B radial trajectory.}
Points are $\rho_{t,q}=\operatorname{top}_3(\lambda_{t,q})\in\mathbb{R}^3$, colored by test accuracy. The radial expansion correlates model performance.}
\label{fig:yale_polar_radial}
    \vskip -0.1in
\end{figure}

\begin{figure}[htbp]
    \vskip -0.1in
\centering
\includegraphics[width=0.7\columnwidth]{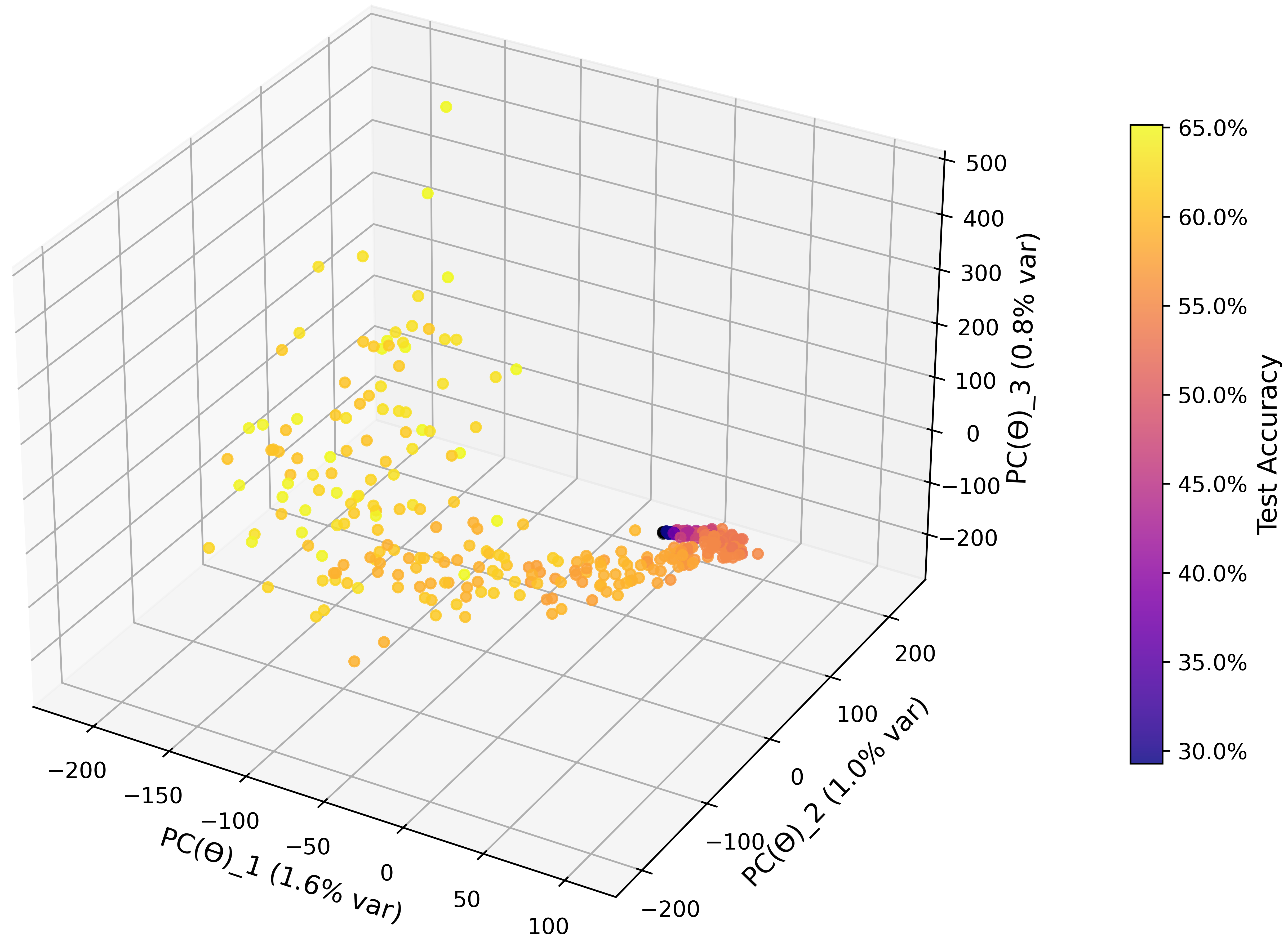}
\caption{\textbf{Yale B angle trajectory.}
$\operatorname{PCA}_3 ( \{ \Theta_t \}_t )$ does not reveal particular patterns in this projected angular coordinate.}
\label{fig:yale_polar_angle_v1}
    \vskip -0.12in
\end{figure}

\begin{enumerate}
    \item The training model trajectory gives radial coordinates $\rho_{t,q}$
    from Eq.~\eqref{E: top k eigenvalues}, expanding outward with increasing
    test accuracy in \textbf{Fig.}~\ref{fig:yale_polar_radial}.  This indicates the spectral coordinates $\lambda_{t,q}$ are strongly correlated with model
    performance.

    \item Direct 3D visualization of the angular coordinate in
    Eq.~\eqref{E: unitary angle} is shown in \textbf{Fig.}~\ref{fig:yale_polar_angle_v1}.
    Here PCA is applied to the high-dimensional Hermitian generators
    $\Theta_t\in\operatorname{Herm}(2^{10})$, denoted by
    $\operatorname{PCA}_3(\{\Theta_t\}_t)$.  This direct angular projection does
    not reveal an obvious pattern.
\end{enumerate} 

\begin{figure}[htbp]
    \vskip -0.2in
\centering
\includegraphics[width=0.8\columnwidth]{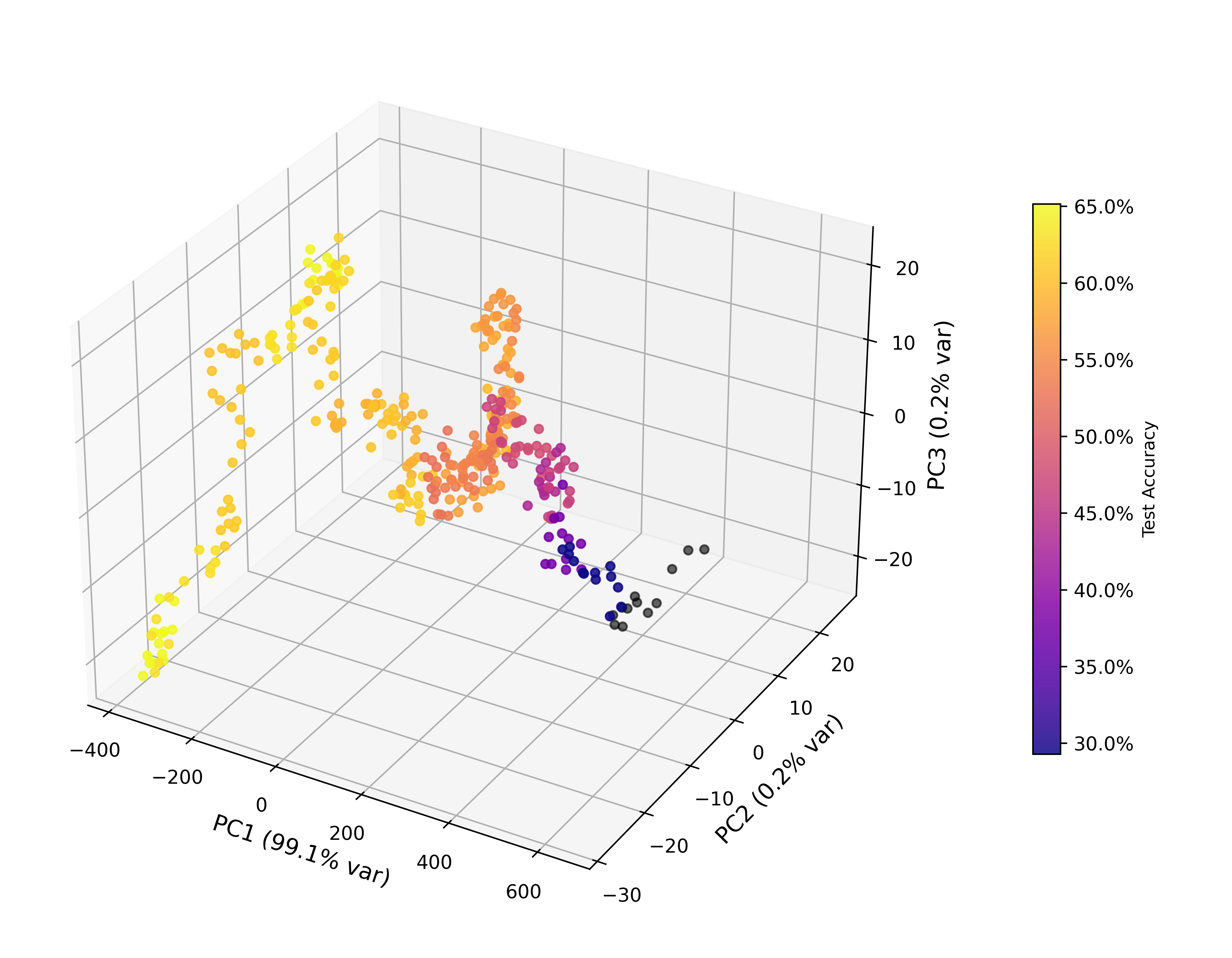}
\caption{\textbf{Yale B eigenphase PCA.}
Extracting $\a_t$ from Eq.~\eqref{E: theta spectral transform} and $\operatorname{PCA}_3(\{\alpha_t\}_t)$ reveals a pattern correlated with model accuracy along PC1 (cf. \textbf{Fig.}~\ref{fig:yale_polar_angle_v1}).}
\label{fig:yale_polar_angle_v0}
    \vskip -0.1in
\end{figure}

To further inspect the angular motion, we pass from $\Theta_t$ to its ordered
eigenphase spectrum. Let $N=2^{10}$ and write
\begin{equation}\label{E: theta spectral transform}
    \Theta_t
    =
    V_t\operatorname{diag}(\alpha_t)V_t^\dagger,
    \quad
    \alpha_t=(\mu_{t,1},\ldots,\mu_{t,N})\in\mathbb{R}^{N},
\end{equation}
where $\mu_{t,1}\leq\cdots\leq\mu_{t,N}$ are counted with multiplicity. Since $U(\theta_t)=e^{i\Theta_t}$, the entries of $\alpha_t$ are the unwrapped eigenphases of the lifted unitary. Applying $\operatorname{PCA}_3(\{\alpha_t\}_t)$ gives \textbf{Fig.}~\ref{fig:yale_polar_angle_v0}, which reveals a strong accuracy-correlated trend along PC1.

The change from \textbf{Fig.}~\ref{fig:yale_polar_angle_v1} to \textbf{Fig.}~\ref{fig:yale_polar_angle_v0} reveals a phenomenon that is not visible in the direct angular projection. In the eigenphase coordinates, the trajectory aligns strongly with the leading PCA component, with PC1 alone accounting for over $99\%$ of the explained variance. Moreover, the model accuracy appears correlated with this hidden eigenphase axis. We do not yet have a complete explanation for this phenomenon; its possible interpretation is discussed further in Sec.~\ref{subsec:polar_interpretation}. Below, we use perturbations and a second dataset to test whether the pattern is robust.

\subsection{Perturbation and cross-dataset comparison}\label{subsec:polar_comparison_exp}

\begin{figure}[htbp]
    \vskip -0.1in
\centering
\includegraphics[width=0.8\columnwidth]{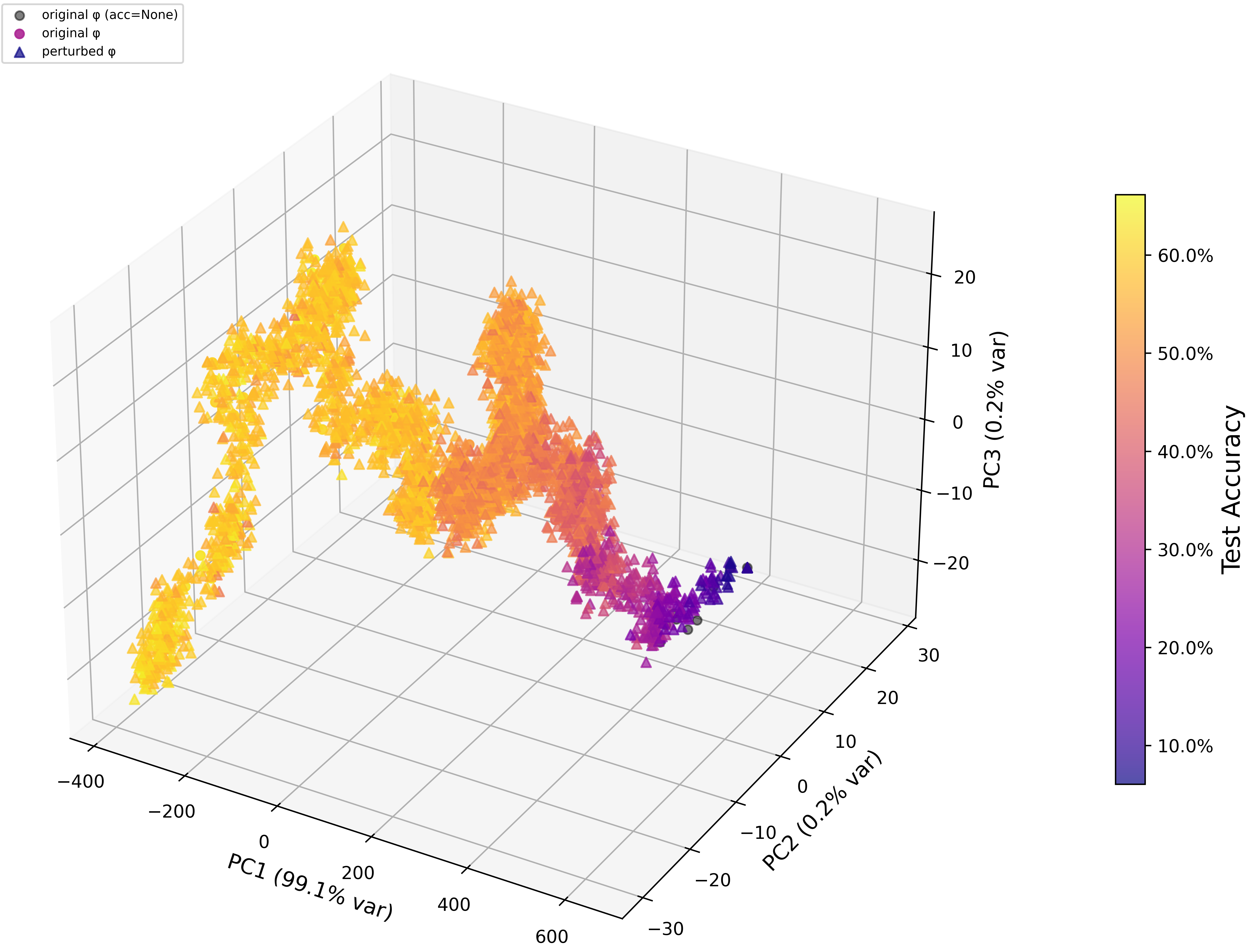}
\caption{\textbf{Yale B eigenphase perturbations.} Circles are trained points $\alpha_t$ from original model $\Theta_t$; triangles are $\widetilde \alpha_t$ from perturbed models $\widetilde\Theta_t=\Theta_t+\delta\Theta_t$. The plot shows $\operatorname{PCA}_3(\{\widetilde\alpha_t\}_t)$ with the original trajectory overlaid (\textbf{Fig.}~\ref{fig:yale_polar_angle_v0}).}
\label{fig:yale_polar_angle_perturbed}
    \vskip -0.1in
\end{figure}

We perturb the training trajectory to form nearby model evolutions $t\mapsto\Theta_t+\delta\Theta_t$. These perturbed models expand the original training path. Extracting the corresponding eigenphase variations $\alpha_t+\delta\alpha_t$ from Eq.~\eqref{E: theta spectral transform} and projecting by $\operatorname{PCA}_3(\{\alpha_t+\delta\alpha_t\}_t)$ gives the widened trail in \textbf{Fig.}~\ref{fig:yale_polar_angle_perturbed}.  This makes the relation between test accuracy and the leading eigenphase component more visible.

\begin{figure}[htbp]
    \vskip -0.15in
\centering
\includegraphics[width=1.0\columnwidth]{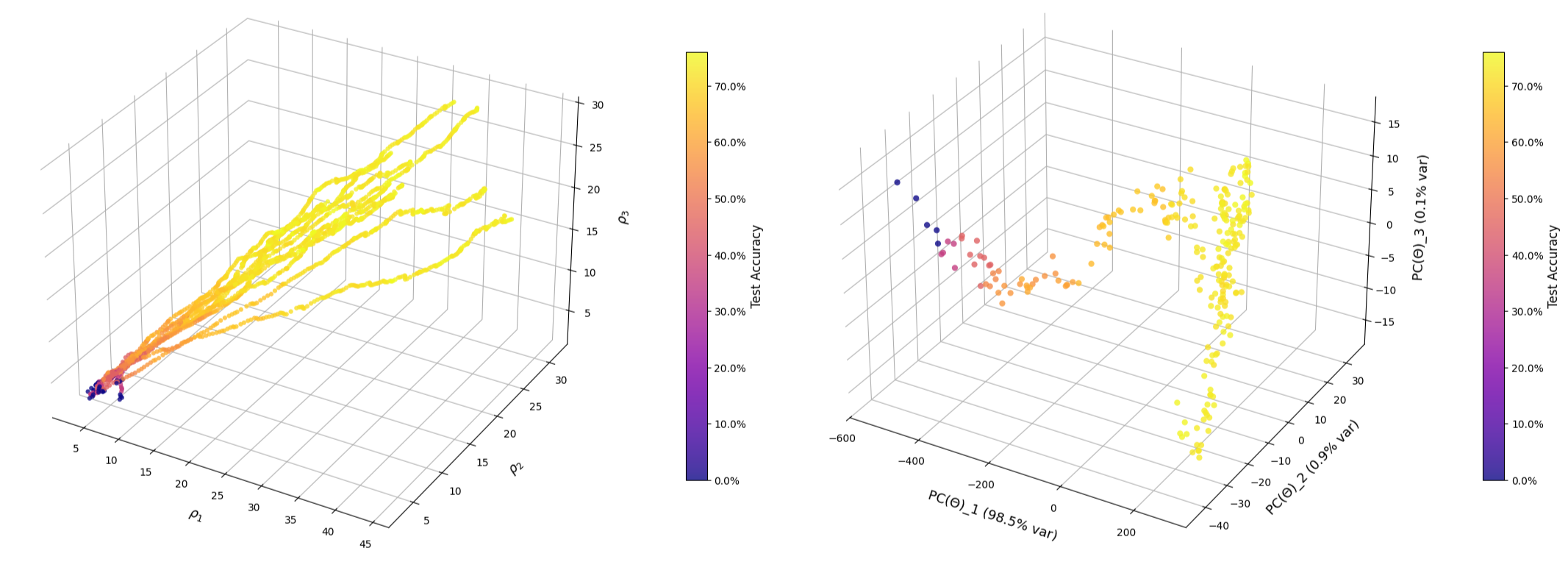}
\caption{\textbf{MNIST Quasi-polar training model trajectories.}
\textbf{[Left]} Radial evolution $\rho_{t,q}$. \textbf{[Right]} PCA eigenphases.}
\label{fig:combined_mnist}
\vskip -0.1in
\end{figure}

The same procedure is then applied to MNIST.  The DANO radial coordinates again show spectral expansion along with test accuracy in \textbf{Fig.}~\ref{fig:combined_mnist}[Left]. The eigenphase trajectory in \textbf{Fig.}~\ref{fig:combined_mnist}[Right] also correlates with the dominant PC1 direction, up to sign, with PC1 alone explaining about $98.5\%$ of the eigenphase variance.

\subsection{Interpretation}\label{subsec:polar_interpretation}

This suggests that the eigenphase variables may capture hidden structure in
$\operatorname{Herm}(2^{10})$ that is relevant to model performance, although
the precise mechanism is not yet clear.  To give a possible interpretation of
the difference between \textbf{Fig.}~\ref{fig:yale_polar_angle_v1} and
\textbf{Fig.}~\ref{fig:yale_polar_angle_v0}, consider the spectral decomposition in
Eq.~\eqref{E: theta spectral transform}.  Under the change of basis $V_t$, write
\[
    \ket{\psi_x'}:=V_t^\dagger\ket{\psi_x}, \quad H_t':=V_t^\dagger H V_t .
\]
Then the final measurement can be expressed as
\[
    \bra{\psi_x}U^\dagger(\theta_t)HU(\theta_t)\ket{\psi_x} = \sum_{a,b=1}^{N} e^{i(\mu_{t,b}-\mu_{t,a})} \overline{\psi'_{x,a}}\, (H'_t)_{ab}\, \psi'_{x,b}.
\]
This expression shows that the phase gaps $\mu_{t,b}-\mu_{t,a}$ affect the interference between eigendirections before measurement. One possible physical picture is that training may partially organize these phase gaps so that different inputs, or different classes, lead to more distinguishable interference patterns. The DANO weights $\lambda_t$ then assign the corresponding spectral measurement weights.

In our experiments, the observed QML dynamics are radial spectral expansion coupled with alignment in the projected eigenphase coordinates. A complete explanation of the eigenphase alignment remains open.

\section{Conclusion}\label{sec_conclusion}

This work used DANO to view variational quantum models through the quasi-polar chart defined in Eqs.~\eqref{E: polar Hermitian exponential} and \eqref{E: radial angular trajectories}. This separates model evolution into radial spectral motion and angular unitary motion.

The experiments show that the radial variables $\lambda_t$ expand with model accuracy, while the full angular coordinates $\Theta_t$ do not show particular structure on a single Lie-algebra direction. The most visible angular structure appears after the spectral transformation of Eq.~\eqref{E: theta spectral transform}, where the eigenphase variables $\alpha_t$ reveal an accuracy-correlated axis. Thus, the observed dynamics are better summarized as radial spectral expansion together with angular eigenphase transformation.

This quasi-polar viewpoint is not limited to QML tasks. Any VQA based on observable expectations can be analyzed via the developed DANO decomposition, providing a tool for tracking the evolution of quantum models.

\bibliographystyle{IEEEtran}
\bibliography{references}

\end{document}